# Substitutions for tilings $\{p, q\}$

M. Margenstern[1] and G. Skordev[2]
[1] Université de Metz, LITA, EA 3097, Île du Saucy,
57045 METZ Cédex, FRANCE, margens@sciences.univ-metz.fr
[2] Universität Bremen, Fachbereit für Mathematik, CeVis Centre,
Universitätsallee, 29, 28359 BREMEN, GERMANY, skordev@cevis.uni-bremen.de

In this paper we consider tiling $\{p, q\}$ of the Euclidean space and of the hyperbolic space, and its dual graph $\Gamma_{q,p}$ from a combinatorial point of view. A substitution $\sigma_{q,p}$ on an appropriate finite alphabet is constructed. The homogeneity of graph $\Gamma_{q,p}$ and its generation function are the basic tools for the construction. The tree associated with substitution $\sigma_{q,p}$ is a spanning tree of graph $\Gamma_{q,p}$. Let $u_n$ be the number of tiles of tiling $\{p, q\}$ of generation $n$. The characteristic polynomial of the transition matrix of substitution $\sigma_{q,p}$ is a characteristic polynomial of a linear recurrence. The sequence $(u_n)_{n\geq 0}$ is a solution of this recurrence. The growth of sequence $(u_n)_{n\geq 0}$ is given by the dominant root of the characteristic polynomial. The result of this paper is related to [6, 5, 7, 8, 9, 10, 2].

## 1 Combinatorial structure of tiling $\{p, q\}$ induced by the generation function

Here we shall develop [2] using different combinatorial method.

We consider tiling $\{p, q\}$ of the Euclidean or the hyperbolic plane by regular $p$-gons such that each of its vertex is shared by $q$ gons, i. e., $(p-2)(q-2) \geq 4$, [11], Ch. 5, 3.3. Denote by $\Gamma_{q,p}$ the dual graph of tiling $\{p, q\}$. Remind that the vertices of $\Gamma_{q,p}$ correspond to the tiles of $\{p, q\}$. Two vertices of $\Gamma_{q,p}$ are connected in this graph iff the corresponding tiles have a common edge. We assume that the dual tiling $\{q, p\}$ and the dual graph $\Gamma_{q,p}$ are naturally imbedded in the plane of tiling $\{p, q\}$. Precisely, the vertices of $\Gamma_{q,p}$ are the centres of the corresponding tiles of $\{p, q\}$. Let $V_{q,p}$ be the set of the vertices of graph $\Gamma_{q,p}$ and let $E_{q,p}$ be the set of its edges. We denote by $e_{v_1,v_2}$ the edge of $\Gamma_{q,p}$, connecting vertices $v_1$ and $v_2$. Every vertex $v \in V_{q,p}$ is connected with $p$ vertices, i.e., the degree of $v$ in $\Gamma_{q,p}$ is $p$. The vertices, connected with $v$, are called neighbors of $v$.

We remind the definition of a metric $d(.,.)$ on set $V_{q,p}$. Let $\gamma = e_{v_1,w_2}e_{w_2,w_3}\dots e_{w_k,v_2}$ be a path in graph $\Gamma_{q,p}$, connecting vertices $v_1$ and $v_2$. We say that the length $l(\gamma)$ of path $\gamma$ is $l(\gamma) = k$. Then the metric is defined by $d(v_1, v_2) = min\{l(\gamma) \ : \ \gamma \text{ is connecting } v_1 \text{ with } v_2\}$.

Choose and fix a vertex $\tilde{v} \in V_{q,p}$. We call this vertex *a root* the graph $\Gamma_{q,p}$.

**Definition 1.1** *The generation function $g : V_{q,p} \longrightarrow \{0, 1, 2, \dots\}$ is defined by $g(v) = d(\tilde{v}, v)$ for $v \in V_{q,p}$. The value $g(v)$ is called the generation of vertex $v$.*

**Remark 1.1** *Every tile of $\{p, q\}$ is an image of the tile corresponding to the root of $\Gamma_{q,p}$ by an iteration of symmetries with respect to lines supporting edges of tiles of $\{p, q\}$. The minimal number of such symmetries is the generation of the tile. The vertices of graph $\Gamma_{q,p}$ correspond to the tiles of tiling $\{p, q\}$ and the generation of a vertex is the generation of the tile, corresponding to it.*

Let $v'$ be a neighbor of $v$. Vertex $v'$ is called *a successor* of $v$ if $g(v') = g(v) + 1$. In this case $v$ is called *predecessor* of $v'$.

Denote by $V_n$ the set of all vertices of generation $n$, i.e., $V_n = \{v \in V_{q,p} \ : \ g(v) = n\}$.

The choice of the root $\tilde{v}$ and the generation function $g(n)$ impose a structure on graph $\Gamma_{q,p}$ and therefore a structure of tiling $\{p,q\}$. Our goal is to describe this structure. We shall find a *substitution (a morphism)* $\sigma_{q,p} : A_{q,p} \longrightarrow A^*_{q,p}$ on an appropriate alphabet $A_{q,p}$. This substitution generates set $V_n$ and the tree associated with $\sigma_{p,q}$ is a spanning tree of graph $\Gamma_{q,p}$ with root $\tilde{v}$. Moreover, the set of vertices of this tree of generation $n$ coincides with set $V_n$. This implies that sequence $(u_n)_{n\geq 0}$, $u_n = card(V_n)$, is a solution of a linear recurrence with constant coefficients. The characteristic polynomial of this recurrence is the characteristic polynomial of the transition matrix of substitution $\sigma_{q,p}$. The homogeneity of graph $\Gamma_{q,p}$ is very important for the analysis of its combinatorial structure. Moreover, it is known from the proof of Poincaré's theorem that the local considerations are consistent, see [11], Ch. 2, , 1.5. For this reason the proofs of many assertions are by a simple induction on the generation of vertices and, consequently, they are skipped. Let us note that the combinatorial structure of $\Gamma_{q,p}$ depends on the parity of $q$.

## 1.1 Orientation, elementary cycles and relative generation in $\Gamma_{q,p}$

**Lemma 1.1** *There are vertices $v_1, v_2 \in V_{2k+1,p}$, connected by an edge $e_{v_1,v_2} \in E_{2k+1,p}$, such that $g(v_1) = g(v_2)$. There are no such vertices in $E_{2k,p}$.*

Proof. The assertion follows by an induction on generation $g(v_1)$ of vertex $v_1$.

**Remark 1.2** *Lemma 1.1 manifests an important difference between graphs $\Gamma_{2k,p}$ and $\Gamma_{2k+1,p}$.*

Let $e_{v_1,v_2} \in E_{q,p}$ and let $g(v_2) = g(v_1) + 1$. We consider edge $e_{v_1,v_2}$ with *the orientation* in the direction from $v_1$ to $v_2$.

**Corollary 1.1** *All edges of graph $\Gamma_{2k,p}$ are oriented, i.e., $\Gamma_{2k,p}$ is a directed graph.*

A closed path $C = e_{v_1,v_2}e_{v_2,v_3}\dots e_{v_q,v_1}$ in $\Gamma_{q,p}$ is called *an elementary cycle*. All elementary cycles of graph $\Gamma_{q,p}$ have a length $q$, and the elementary cycles are boundaries of the tiles of the dual tiling $\{q,p\}$ of tiling $\{p,q\}$

**Definition 1.2** *Let $g(C) = min\{g(v_i) \ : \ v_i \in V_{q,p}, \ v_i \in C\}$.*

We call $g(C)$ *a generation* of elementary cycle $C$.

**Lemma 1.2** *Let $C$ be an elementary cycle of $\Gamma_{2k,p}$. Then the set*

$$\{v_0, v_{1l}, v_{1r}, \dots, v_{(k-1)l}, v_{(k-1)r}, v_k\}$$

*of the vertices of $C$ satisfies:*

1. *$g(v_0) = g(C)$;*
2. *$g(v_{il}) = g(v_{ir}) = g(v_0) + i, \ i = 1, 2, \dots, k-1$;*
3. *$g(v_k) = g(v_0) + k$.*

Proof. By induction on $g(C)$.

An elementary cycle of $\Gamma_{2k,p}$ is represented on Fig. 1(left).

**Lemma 1.3** *Let $C$ be an elementary cycle of $\Gamma_{2k+1,p}$. Two cases are possible:*

1. *The set*
$$\{v_{0l}, v_{0r}, v_{1l}, v_{1r}, \ldots, v_{(k-1)l}, v_{(k-1)r}, v_k\}$$
*of the vertices of $C$ satisfies: $g(v_{il}) = g(v_{ir}) = g(C) + i,\ i = 0, 1, 2, \ldots, k-1,$ and $g(v_k) = g(C) + k$;*

2. *The set*
$$\{v_0, v_{1L}, v_{1R}, \ldots, v_{(k-1)L}, v_{(k-1)R}, v_{kL}, v_{kR}\}$$
*of the vertices of $C$ satisfies: $g(v_0) = g(C),\ g(v_{iL}) = g(v_{iR}) = g(C) + i,\ i = 1, 2, \ldots, k.$*

Proof. By induction on $g(C)$.

We shall call the elementary cycles of Lemma 1.3.1 *elementary cycles of the first kind.* The elementary cycles of Lemma 1.3.2 are called *elementary cycles of the second kind.* Both of them are represented on Fig. 1(midle, right).

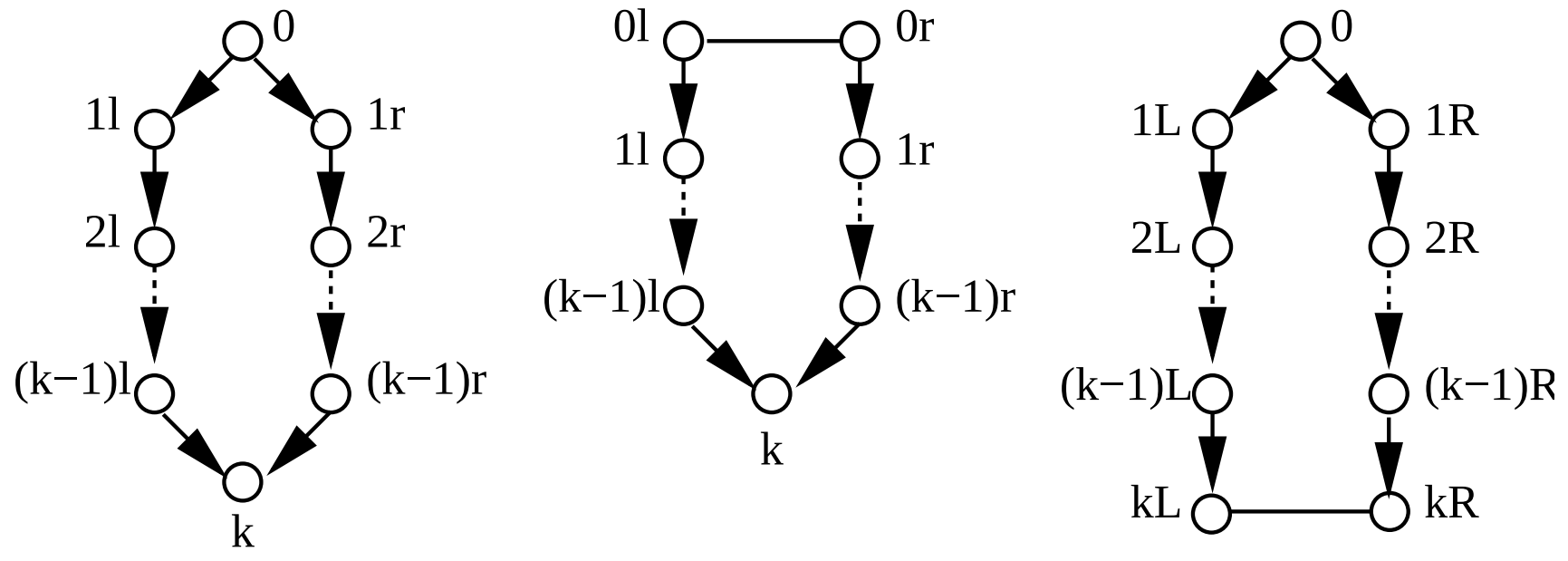

Fig. 1. *Elementary cycle of graph $\Gamma_{2k,p}$, Lemma 1.2(left) and elementary cycles of $\Gamma_{2k+1,p}$, Lemma 1.3(midle, right).*

**Remark 1.3** *Observe that edges $e_{v_{0l},v_{0r}}$ and $e_{v_{kL},v_{kR}}$ of the elementary cycles on Fig. 1(midle, right) are not oriented.*

**Definition 1.3** *Let $C$ be an elementary cycle of $\Gamma_{q,p}$ and let $v$ be a vertex of $C$.The relative generation $g_C(v)$ of $v$ in $C$ is defined as follows:*

1. ***<u>Case q=2k</u>****:*
$$C = \{v_0, v_{1l}, v_{1r}, \ldots, v_{(k-1)l}, v_{(k-1)r}, v_k\},$$
*here we use the notation of Lemma 1.2.*

   *Then $g_C(v_0) = 0,\ g_C(v_{il}) = il,\ g_C(v_{ir}) = ir$ for $i = 1, \ldots, k-1,$ and $g_C(v_k) = k$;*

2. ***Case q=2k+1, elementary cycle of the first kind**:*

$$C = \{v_{0l}, v_{0r}, \dots, v_{(k-1)l}, v_{(k-1)r}, v_k\},$$

*here we use the notations of Lemma 1.3(1).*

*Then $g_C(v_{il}) = il,\ g_C(v_{ir}) = ir$ for $i = 0, \dots, k-1$, and $g_C(v_k) = k$;*

3. ***Case q=2k+1, elementary cycle of the second type:***

$$C = \{v_0, v_{1L}, v_{1R}, \dots, v_{(k-1)L}, v_{(k-1)R}, v_{kL}, v_{kR}\},$$

*here we use the notations of Lemma 1.3(2).*

*Then $g_C(v_0) = 0,\ g_C(v_{iL}) = iL,\ g_C(v_{iR}) = iR$ for $i = 1, \dots, k$.*

## 1.2 Types of the vertices of $\Gamma_{q,p}$

Here we shall define a type $t(v)$ of a vertex $v \in V_{q,p}$. For this we consider two cases $q = 2k$ and $q = 2k + 1$.

**Case q=2k**

Let $A_{2k,p} = \{0, 1l, 1r, \dots, (k-1)l, (k-1)r, k\}$. We shall define the map $t : V_{2k,p} \longrightarrow A^p_{2k,p}$. For this we need some preliminary considerations.

Let $v \in V_{2k,p}$ and $C(v) = \{C_1, \dots, C_p\}$ be the set of all elementary cycles in $\Gamma_{2k,p}$, which share $v$ as a common vertex. We shall define an order of set $C(v)$. For this we define the first element of this set and use the positive orientation of the plane (Euclidean or hyperbolic) where tiling $\{p, 2k\}$ and its dual graph $\Gamma_{2k,p}$ are embedded.

**Lemma 1.4** *Let $v \in \Gamma_{2k,p},\ k \geq 2,\ p \geq 4$ and let $m(v) = min\{g(C)\ :\ C \in C(v)\}$. The following cases are possible:*

1. *There is only one elementary cycle $C_i \in C(v)$ with $m(v) = g(C_i)$;*
2. *There are only two elementary cycles $C_s, C_t \in C(v)$ with $m(v) = g(C_s) = g(C_t)$, and $g_{C_s}(v) = 1l,\ g_{C_t}(v) = 1r$;*
3. *$g(C_1) = \cdots = g(C_p)$ for $C_i \in C(v),\ i = 1, \dots, p$, and $v = \tilde{v}$.*

Proof. By induction on $g(v)$, see Fig. 2.

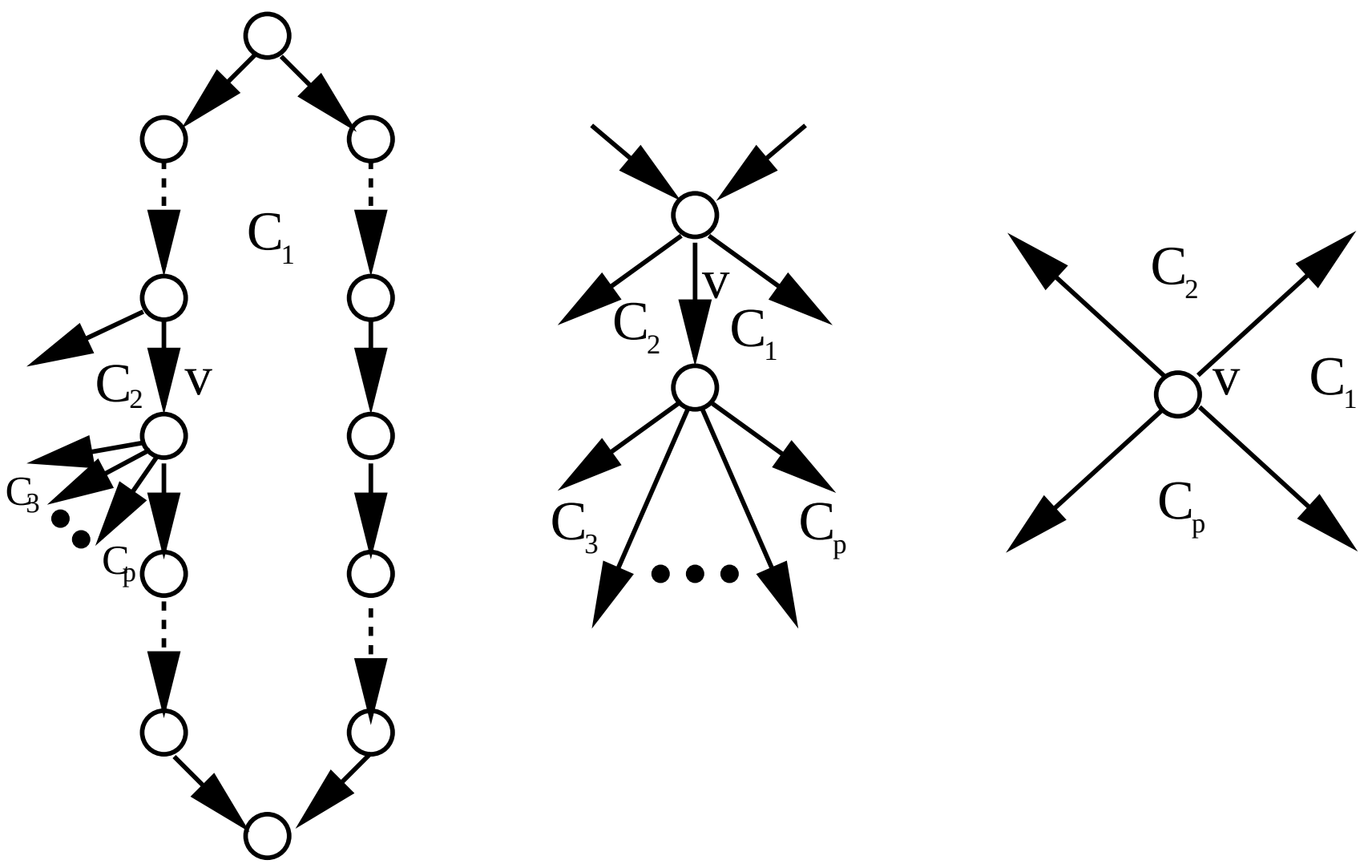

Fig. 2. *Elementary cycles of Lemma 1.4(1)(left), (2)(midle), (3)(right).*

**Definition 1.4** *The first element of set $C(v)$, $v \in V_{2k,p}$ is defined as follows:*

1. *It is the unique element $C_i \in C(v)$ in the case of Lemma 1.4(1);*
2. *It is element $C_s \in C(v)$ in the case of Lemma 1.4(2);*
3. *It is an arbitrary element of $C(v)$ in the case of Lemma 1.4(3).*

**Lemma 1.5** *Let $v \in \Gamma_{2k,3}$, $k \geq 3$ and let $m(v) = min\{g(C) \ : \ C \in C(v)\}$. The following cases are possible:*

1. *There is only one elementary cycle $C_i \in C(v)$ with $m(v) = g(C_i)$;*
2. *There are only two elementary cycles $C_s, C_t \in C(v)$ with $m(v) = g(C_s) = g(C_t)$, and one of the following cases hold*
   (a) $g_{C_s}(v) = 1l$, $g_{C_t}(v) = 1r$;
   (b) $g_{C_s}(v) = 2l$, $g_{C_t}(v) = 2r$;
3. *$g(C_1) = g(C_2) = g(C_3)$ for $C_i \in C(v)$, $i = 1, 2, 3$, and $v = \tilde{v}$.*

Proof. By induction on $g(v)$ and Lemma 1.3, see Fig. 3.

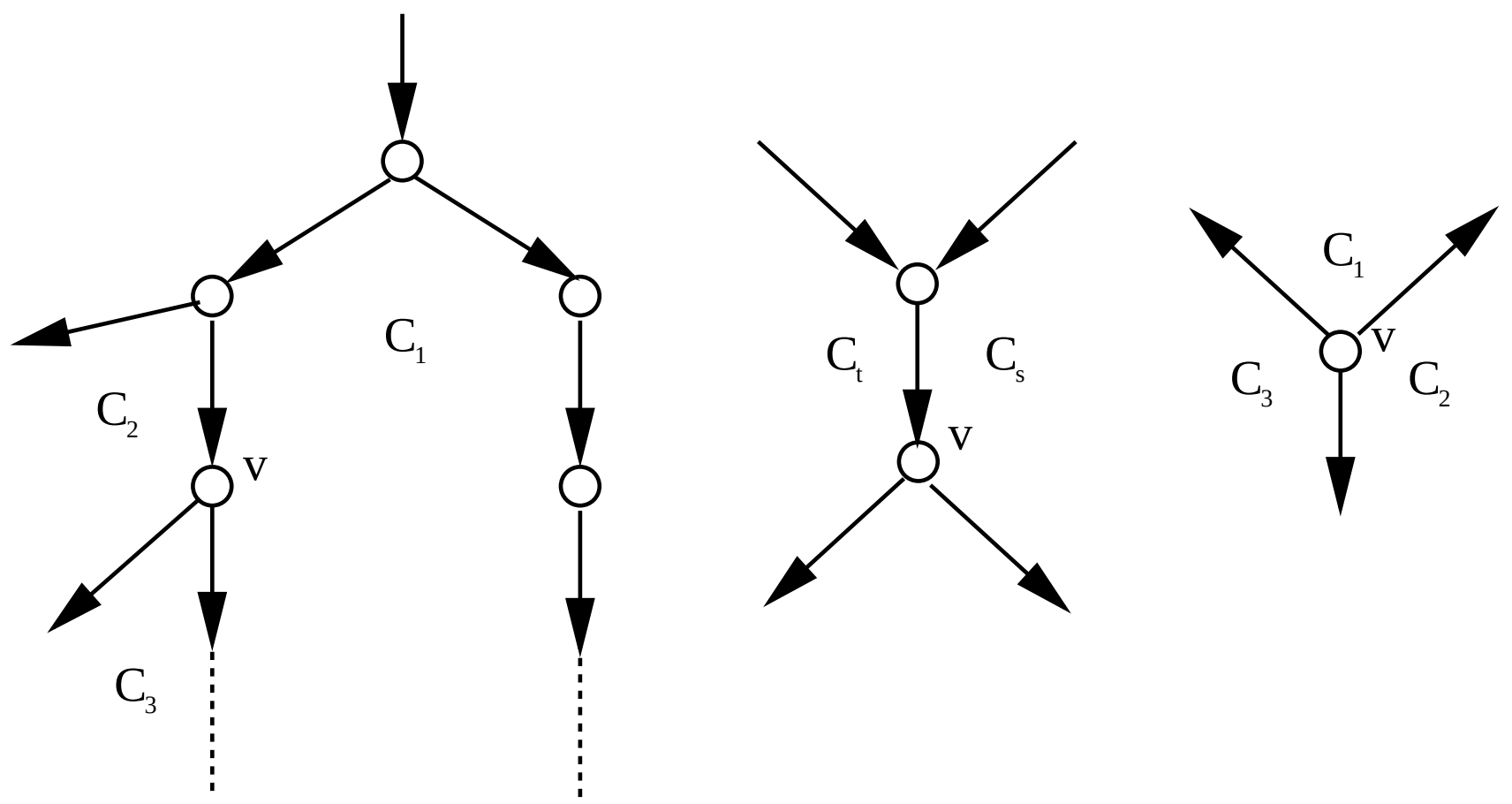

Fig. 3. *Elementary cycles of Lemma 1.5(1)(left), (2a)(midle), (3)(right).*

**Definition 1.5** *The first element of set $C(v),\ v \in V_{2k,3},\ k \geq 3$ is defined as follows:*

1. *It is the unique element $C_i \in C(v)$ in the case of Lemma 1.5(1);*
2. *It is element $C_s \in C(v)$ in the case of Lemma 1.4(2);*
3. *It is an arbitrary element of $C(v)$ in the case of Lemma 1.5(3).*

We denote the first element by $C_1$ in all cases.

The order $C_1 < \cdots < C_p$ of set $C(v) = \{C_1, \ldots, C_p\}$ is defined by the condition that chain $C_1 \ldots C_p$ is in a positive direction in the plane of tiling $\{p, q\}$.

**Definition 1.6** *Let $v \in V_{2k,p}$. The type $t(v)$ of vertex $v$ is defined as $t(v) = g_{C_1}(v) \cdots g_{C_p}(v)$, where $C(v) = \{C_1, \ldots, C_p\}$ and $C_1 < \cdots < C_p$.*

Observe that $t(v) \in A^p_{2k,p}$, where $A_{2k,p} = \{0, 1l, 1r, \ldots, (k-1)l, (k-1)r, k\}$.

**Lemma 1.6** *Let $v \in V_{2k,p},\ k \geq 2,\ p \geq 4$. The following types $t(v)$ are possible:*

1. *$a_i = (il)(1r)0^{p-2},\ i = 1, \ldots, k-1;$*
2. *$a_k = k(1r)0^{p-3}(1l);$*
3. *$\bar{a}_j = (jr)0^{p-2}(1l),\ 2 \leq j \leq k-1;$*
4. *$a_0 = 0^p$.*

*Moreover, $card\{t(v)\ :\ v \in V_{2k,p}\} = 2k - 1 = q - 1$.*

Proof. By induction on $g(v)$ and Lemma 1.4, see Fig. 2.

**Lemma 1.7** *Let $v \in V_{2k,3},\ k \geq 3$. The following types $t(v)$ are possible:*

1. *$a_i = (il)(1r)0,\ i = 1, \ldots, k-1;$*

2. $a_k = k(1r)(1l)$;

3. $a_{k+1} = (2l)(2r)0$;

4. $\bar{a}_j = (jr)0(1l),\ j = 2, \dots, k-1$;

5. $a_0 = 0^3$.

Proof. By induction on $g(v)$. See Fig. 3.

<u>**Case q=2k+1**</u>

Let

$$A_{2k+1,p} = \{0l, 0r, \dots, (k-1)l, (k-1)r, k, 0, 1L, 1R \dots, kL, kR\}.$$

We shall define a map $t : V_{2k+1,p} \longrightarrow A^p_{2k+1,p}$. For this we follow the procedure used in the previous case $q = 2k$.

Let $C(v) = \{C_1, \dots, C_p\}$, where $v$ is a vertex with $v \in C_i,\ i = 1, \dots, p$. We shall define an order $C_1 < \dots < C_p$ of set $C(v)$. Then $t(v)$ is defined as $t(v) = g_{C_1}(v) \dots g_{C_p}(v)$. To define the order of $C(v)$ we define its first element and use the positive orientation of the plane of tiling $\{p, 2k+1\}$.

**Lemma 1.8** *Let $v \in V_{2k+1,p},\ k \geq 2,\ p \geq 4$ and let $m(v) = min\{g(C)\ :\ C \in C(v)\}$. The following cases are possible:*

1. *There is only one elementary cycle $C_i \in C(v)$ with $m(v) = g(C_i)$;*

2. *There are only two elementary cycles $C_s, C_t \in C(v)$ with $m(v) = g(C_s) = g(C_t)$,and one of the following cases hold:*

   (a) $g_{C_s}(v) = 1l,\ g_{C_t}(v) = 1R$;

   (b) $g_{C_s}(v) = 1r,\ g_{C_t}(v) = 1L$;

   (c) $g_{C_s}(v) = 1L,\ g_{C_t}(v) = 1R$;

3. $g(C_1) = \dots = g(C_p)$ *for* $C_i \in C(v),\ i = 1, \dots, p$.

Proof. By induction on $g(v)$ and Lemma 1.3, see Fig. 4a and Fig. 4b.

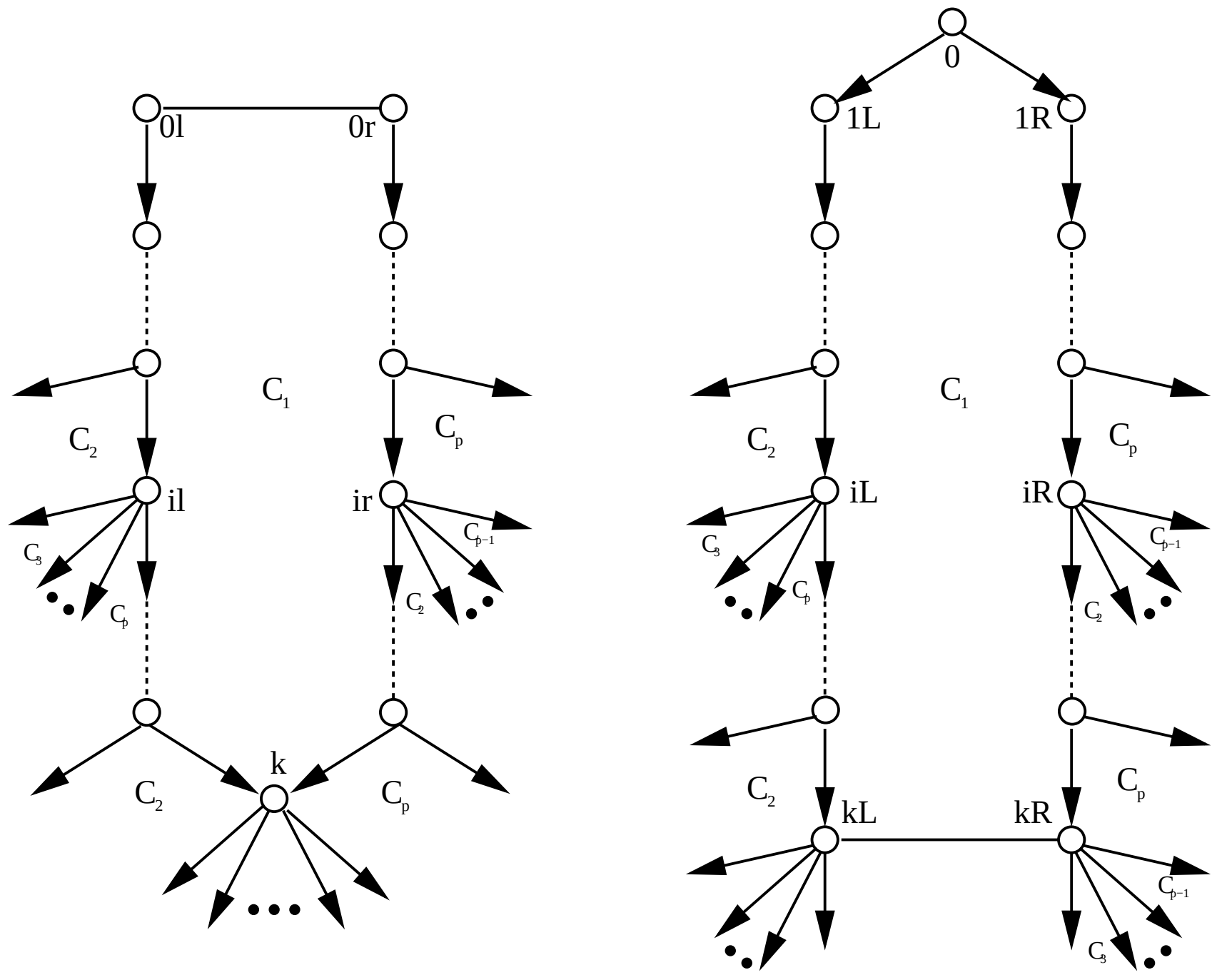

Fig. 4a. Elementary cycles, Lemma 1.8(1).

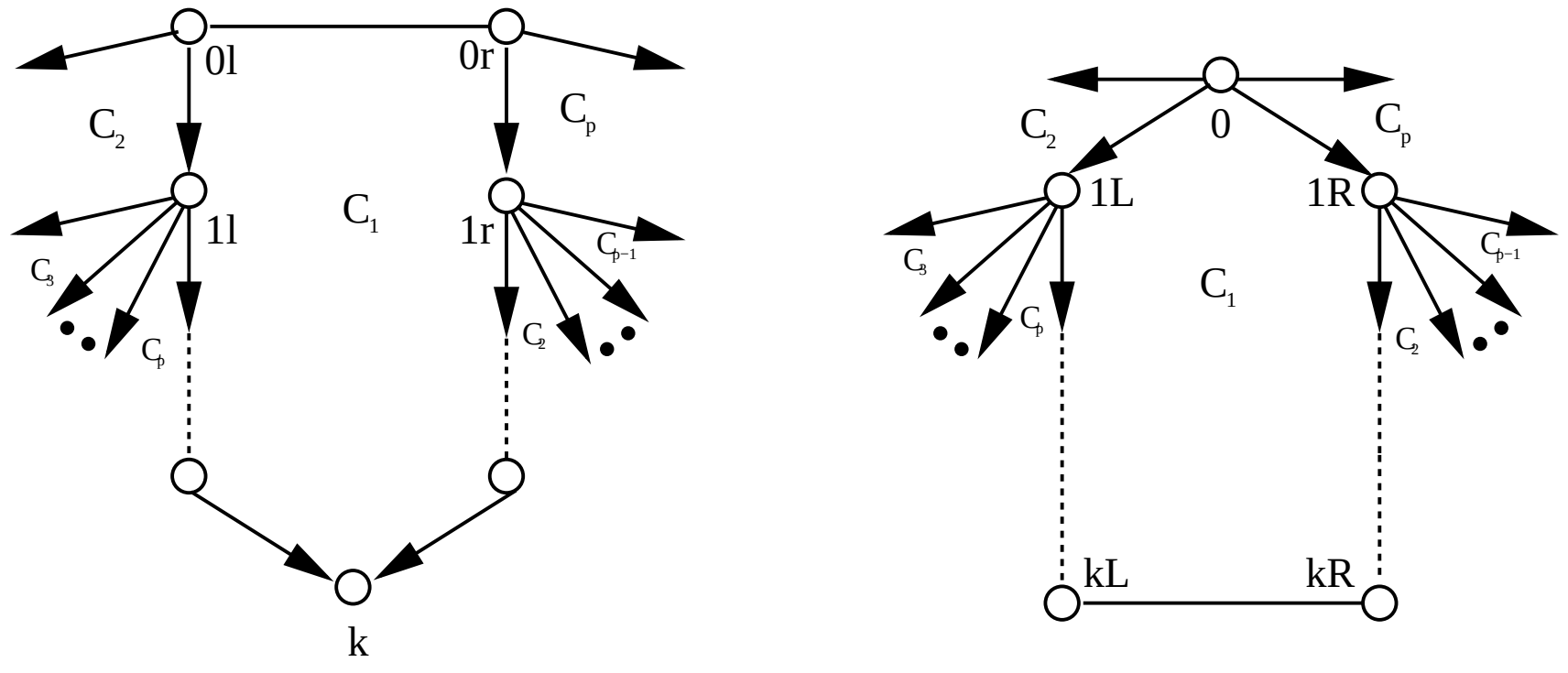

Fig. 4b. Elementary cycles, Lemma 1.8(2).

**Definition 1.7** *The first element of set $C(v)$ is defined as follows:*

1. *$C_i$ in the case of Lemma 1.8(1);*
2. *$C_s$ in the case of Lemma 1.8(2);*
3. *It is an arbitrary element of $C(v)$ in the case of Lemma 1.8(3).*

**Lemma 1.9** *Let $v \in V_{2k+1,3}$, $k \geq 3$ and let $m(v) = min\{g(C) \; : \; C \in C(v)\}$. The following cases are possible:*

1. *There is only one elementary cycle $C_i \in C(v)$ with $m(v) = g(C_i)$;*

2. *There are only two elementary cycles $C_s, C_t \in C(v)$ with $m(v) = g(C_s) = g(C_t)$,and one of the following cases hold:*

   (a) $g_{C_s}(v) = 1l,\ g_{C_t}(v) = 1R$ ;

   (b) $g_{C_s}(v) = 1r,\ g_{C_t}(v) = 1L$ ;

   (c) $g_{C_s}(v) = 1L,\ g_{C_t}(v) = 1R$;

   (d) $g_{C_s}(v) = 2L,\ g_{C_t}(v) = 2R$;

3. $g(C_1) = \cdots = g(C_p)$ *for* $C_i \in C(v),\ i = 1, \ldots, p$.

Proof. By induction on $g(v)$ and Lemma 1.3 see Fig. 5.

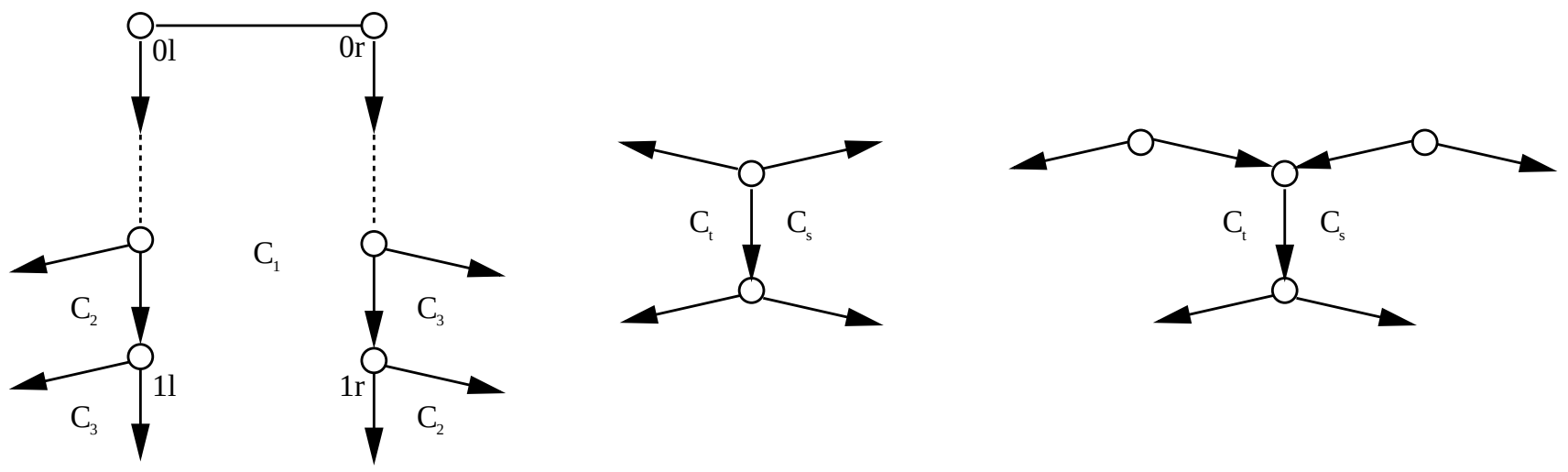

Fig. 5. Elementary cycles, Lemma 1.9(1)(left), Lemma 1.9(2a)(midle), Lemma 1.9(2b)(right).

**Definition 1.8** *The first element of set $C(v)$ is defined as follows:*

1. *$C_i$ in the case of Lemma 1.9(1);*

2. *$C_s$ in the case of Lemma 1.9(2);*

3. *It is an arbitrary element of $C(v)$ in the case of Lemma 1.9(3).*

**Lemma 1.10** *Let $v \in V_{3,p},\ p \geq 6$ and let $m(v) = min\{g(C)\ :\ C \in C(v)\}$. The following cases are possible:*

1. *There are only two elementary cycles $C_s, C_t \in C(v)$ with $m(v) = g(C_s) = g(C_t)$, and $g_{C_s}(v) = 1L,\ g_{C_t}(v) = 1R$ ;*

2. *There are 3 elementary cycles $C_s, C_t, C_u \in C(v)$ with $m(v) = g(C_s) = g(C_t) = g(C_u)$, and $g_{C_s}(v) = 1L,\ g_{C_t}(v) = 1, g_{C_u}(v) = 1R$ ;*

3. $g(C_1) = \cdots = g(C_p)$ *for* $C_i \in C(v),\ i = 1, \ldots, p$.

Proof. By induction on $g(v)$, see Fig. 6.

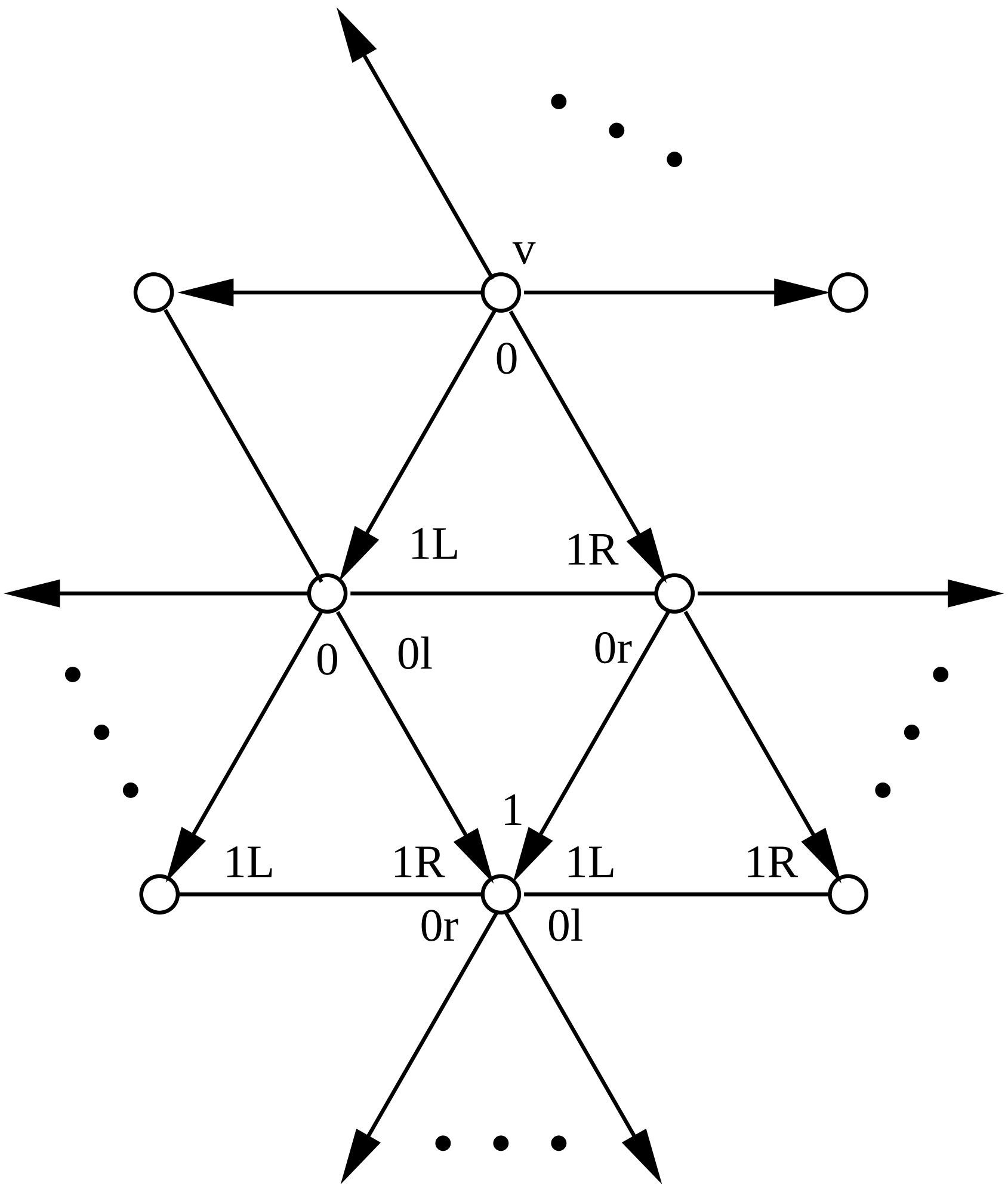

Fig. 6. Elementary cycles, Lemma 1.10.

**Definition 1.9** *The first element of set $C(v)$ is defined as follows:*

1. *$C_i$ in the case of Lemma 1.10(1);*

2. *$C_s$ in the case of Lemma 1.10(2);*

3. *It is an arbitrary element of $C(v)$ in the case of Lemma 1.10(3).*

We denote the first element of $C(v)$ by $C_1$ in all this cases. *The order $C_1 < \dots C_p$ of $C(v) = \{C_1, \dots, C_p\}$* is defined by the condition that chain $C_1 \dots C_p$ is in the positive direction in the plane of tiling $\{p, 2k+1\}$.

**Lemma 1.11** *Let $v \in V_{2k+1,p},\ k \geq 2,\ p \geq 4$. The following types $t(v)$ are possible:*

1. *$a_i = (il)(1R)0^{p-2},\ i = 1, \dots, k-1$;*

2. *$a_k = k(1R)0^{p-3}(1L)$;*

3. *$b_i = (iL)(1R)0^{p-2},\ i = 1, \dots, k-1$;*

4. $b_k = (kL)(1R)0^{p-3}(0l)$;

5. $\bar{a}_j = (jr)0^{p-2}(1L),\ 1 \le j \le k-1$;

6. $\bar{b}_j = (jR)0^{p-2}(1L),\ 2 \le j \le k-1$;

7. $\bar{b}_k = (kR)(0r)0^{p-3}(1L)$;

8. $a_0 = 0^p$.

Proof. Induction on $g(v)$ and Lemma 1.8, see Fig. 4.

**Lemma 1.12** *Let $v \in V_{2k+1,3},\ k \ge 3$. The following types $t(v)$ are possible:*

1. $a_1 = (2R)0(1l)$;

2. $a_i = (il)(1R)0,\ i = 2, \dots, k-1$;

3. $a_k = k(1R)(1L)$;

4. $a_{k+1} = (2R)(2L)0$

5. $b_i = (iL)(1R)0,\ i = 1, \dots, k-1$;

6. $b_k = (kL)(1R)(0l)$;

7. $\bar{a_1} = (2L)(1r)0$;

8. $\bar{a}_j = (jr)0(1L),\ 2 \le j \le k-1$;

9. $\bar{b}_j = (jR)0(1L),\ 2 \le j \le k-1$;

10. $\bar{b}_k = (kR)(0r)(1L)$;

11. $a_0 = 0^3$.

Moreover, $card\{t(v)\ :\ v \in V_{2k+1,p},\ k > 1,\ p \ge 4\} = 4k-1 = 2q-3$.
Proof. By induction on $g(v)$ and Lemma 1.9, see Fig. 5.

**Remark 1.4** *For the sake of simplicity of the notations we take $\bar{a}_1 = (1r)0^{p-2}(1L)$ instead of $(1L)(1r)0^{p-2}$.*

**Lemma 1.13** *Let $v \in V_{3,p},\ k = 1,\ p \ge 6$. The following types $t(v)$ are possible:*

1. $a = (1L)(1R)(0r)0^{p-4}(0l)$;

2. $b = (1L)1(1R)(0r)0^{p-5}(0l)$.

3. $a_0 = 0^p$.

Proof. By induction on $g(v)$ and Lemma 1.10, see Fig. 6.

### 1.3 Types of the successors of the vertices

**Lemma 1.14** *Let $v \in V_{2k,p},\ k \geq 2, p \geq 4$. Let $v_1, \ldots, v_l$ be the successors of vertex $v$. The types of $v_1, \ldots, v_l$ are given by $t(v) \to t(v_1) \ldots t(v_l)$:*

1. $t(v) = a_i \to \bar{a}_2(a_1)^{p-3}a_{i+1},\ 1 \leq i \leq k-1;$
2. $t(v) = a_k \to \bar{a}_2(a_1)^{p-4}a_2;$
3. $t(v) = \bar{a}_j \to \bar{a}_{j+1}(a_1)^{p-3}a_2,\ 2 \leq j \leq k-2;$
4. $t(v) = \bar{a}_{k-1} \to a_k(a_1)^{p-3}a_2;$
5. $t(v) = a_0 \to (a_1)^p.$

Proof. The assertion follows from Lemma 1.6, use Fig. 2.

**Lemma 1.15** *Let $v \in V_{2k,3},\ k \geq 3$. Let $v_1, \ldots, v_l$ be the successors of vertex $v$. The types of $v_1, \ldots, v_l$ are given by $t(v) \to t(v_1) \ldots t(v_k)$:*

1. $t(v) = a_i \to \bar{a}_2 a_{i+1},\ 1 \leq i \leq k-1;$
2. $t(v) = a_k \to a_{k+1};$
3. $t(v) = a_{k+1} \to \bar{a}_3 a_3;$
4. $t(v) = \bar{a}_j \to \bar{a}_{j+1}a_2,\ 2 \leq j \leq k-1;$
5. $t(v) = a_0 \to (a_1)^p.$

Proof. The assertion follows from Lemma 1.7, use Fig. 3.

**Lemma 1.16** *Let $v \in V_{2k+1,p},\ k \geq 2,\ p \geq 4$. Let $v_1, \ldots, v_l$ be the successors of vertex $v$. The types of $v_1, \ldots, v_l$ are given by $t(v) \to t(v_1) \ldots t(v_l)$:*

1. $t(v) = a_i \to \bar{b}_2(b_1)^{p-3}a_{i+1},\ 1 \leq i \leq k-1;$
2. $t(v) = a_k \to \bar{b}_2(b_1)^{p-4}b_2;$
3. $t(v) = \bar{a}_i \to \bar{a}_{i+1}(b_1)^{p-3}b_2,\ 1 \leq j \leq k-2;$
4. $t(v) = \bar{a}_{k-1} \to a_k(b_1)^{p-3}b_2;$

*5.* $t(v) = b_i \to \bar{b}_2(b_1)^{p-3}b_{i+1},\ 1 \le i \le k-1$*;*

*6.* $t(v) = b_k \to \bar{b}_2(b_1)^{p-4}a_1$*;*

*7.* $t(v) = \bar{b}_j \to \bar{b}_{i+1}(b_1)^{p-3}b_2,\ 2 \le j \le k-2$*;*

*8.* $t(v) = \bar{b}_{k-1} \longrightarrow b_k(b_1)^{p-3}b_2$*;*

*9.* $t(v) = \bar{b}_k \to \bar{a}_1(b_1)^{p-4}b_2$*;*

*10.* $t(v) = a_0 \to (b_1)^p$*.*

Proof. The assertion follows from Lemma 1.11, use Fig. 4a, 4b.

**Remark 1.5** *Observe that* $b_k$ *and* $\bar{b}_k$ *are connected in* $\Gamma_{q,p}$ *by a non oriented edge.*

**Lemma 1.17** *Let* $v \in V_{2k+1,3},\ k \ge 3$*. Let* $v_1, \ldots, v_l$ *be the successors of vertex* $v$*. The types of* $v_1, \ldots, v_l$ *are given by* $t(v) \to t(v_1) \ldots t(v_l)$*:*

*1.* $t(v) = a_1 \to \bar{b}_3 a_2$*;*

*2.* $t(v) = a_i \to \bar{b}_2 a_{i+1},\ i = 2, \ldots, k-1$*;*

*3.* $t(v) = a_k \to a_{k+1}$*;*

*4.* $t(v) = a_{k+1} \to \bar{b}_3 b_3$*;*

*5.* $t(v) = \bar{a}_1 \to \bar{a}_2 b_3$*;*

*6.* $t(v) = \bar{a}_i \longrightarrow \bar{a}_{i+1} b_2,\ i = 2, \ldots, k-1$*;*

*7.* $t(v) = b_i \to \bar{b}_2 b_{i+1},\ 1 \le i \le k-1$*;*

*8.* $t(v) = b_k \to a_1$*;*

*9.* $t(v) = \bar{b}_i \to \bar{b}_{i+1} b_2,\ i = 2, \ldots, k-1$*;*

*10.* $t(v) = \bar{b}_k \to \bar{a}_1$*;*

*11.* $t(v) = a_0 \to (b_1)^p$.

Proof. The assertion follows from Lemma 1.12 and Lemma 1.9, use Fig. 5.

**Lemma 1.18** *Let $v \in V_{3,p}, p \geq 6$. Let $v_1, \ldots, v_l$ be the successors of vertex $v$. The types of $v_1, \ldots, v_l$ are given by $t(v) \to t(v_1) \ldots t(v_l)$:*

1. $t(v) = a \to ba^{p-5}b$;

2. $t(v) = b \to ba^{p-6}b$;

3. $t(v) = a_0 \to (b)^p$.

Proof. The assertion follows from Lemma 1.13, use Fig. 6.

## 1.4 Substitution generating a spanning tree of $\Gamma_{q,p}$

Let $\Omega_{q,p} = \{t(v) \ : \ v \in V_{q,p}\}$. Here we shall define a substitution (morphism) $\sigma_{q,p} : \Omega_{q,p} \longrightarrow \Omega^*_{q,p}$.

The $n$-th iteration $\sigma^n_{q,p}(0^p)$ of $0^p \in \Omega_{q,p}$ is a concatenation of words belonging to the set $\{\sigma_{q,p}(w) \ : \ w \in \Omega_{q,p}\}$.

The tree $T_{q,p}$ associated with the orbit $(\sigma^n_{q,p}(0^p))_{n\geq 0}$ of word $0^p \in \Omega_{q,p}$ is a spanning tree of graph $\Gamma_{q,p}$. Remind that tree $T_{q,p}$ is defined as follows. Its vertices are the points of orbit $(\sigma^n_{q,p}(0^p))_{n\geq 0}$. The root of $T_{q,p}$ is $0^p$. Let $\omega \in T_{q,p}$, then the successors of $\omega$ in tree $T_{q,p}$ are all words belonging to $\sigma_{q,p}(\omega)$.

The embedding of tree $T_{q,p}$ in graph $\Gamma_{q,p}$ is defined by induction on the generation of the vertices. The root $0^p$ of $T_{q,p}$ is identified with the root $\tilde{v}$ of $\Gamma_{q,p}$. Assume that the embedding is defined for all vertices of $T_{q,p}$ of generation $\leq k$, and all edges connecting them. Let $\omega \in T_{q,p}, v \in \Gamma_{q,p}$ and let $\omega, v$ be of the same generation $k$. Identify the words in $\sigma_{q,p}(\omega)$ with the successors of $v$ of the same type in the direction of the positive orientation of the plane.

We consider two cases: $q = 2k$ and $q = 2k + 1$ for the definition of substitution $\sigma_{q,p}$.

**Case** $q = 2k, \ k \geq 3, \ p \geq 4$

In this case

$$\Omega_{2k,p} = A^p_{2k,p} = \{a_0, a_1, \ldots, a_k, \bar{a}_2, \bar{a}_3, \ldots, \bar{a}_{k-1}\}.$$

Here we use the notations of Lemma 1.6 and Lemma 1.14.

**Definition 1.10** *Let $q = 2k, k \geq 3, \ p \geq 4$. The substitution $\sigma_{2k,p} : \Omega_{2k,p} \longrightarrow \Omega^*_{2k,p}$ is defined as*

*follows.*

$$\sigma_{2k,p}(a_0) = (a_1)^p;$$

$$\sigma_{2k,p}(a_i) = \bar{a}_2(a_1)^{p-3}a_{i+1},\ 1 \leq i \leq k-2;$$

$$\sigma_{2k,p}(a_{k-1}) = \bar{a}_2(a_1)^{p-3};$$

$$\sigma_{2k,p}(a_k) = \bar{a}_2(a_1)^{p-4}a_2;$$

$$\sigma_{2k,p}(\bar{a}_i) = \bar{a}_{j+1}(a_1)^{p-3}a_2,\ 2 \leq i \leq k-2;$$

$$\sigma_{2k,p}(\bar{a}_{k-1}) = a_k(a_1)^{p-3}a_2.$$

**Case** $q = 4, p = 4$

In this case

$$\Omega_{4,4} = A^4_{4,4} = \{a_0, a_1, a_2\}.$$

**Definition 1.11** *Let* $q = 4, p = 4$. *The substitution* $\sigma_{4,4} : \Omega_{4,4} \longrightarrow \Omega^*_{4,4}$ *is defined as follows.*

$$\sigma_{4,4}(a_0) = (a_1)^4;$$

$$\sigma_{4,4}(a_1) = a_2a_1;$$

$$\sigma_{4,4}(a_2) = a_2.$$

**Case,** $q = 2k, k \geq 4, p = 3$

In this case

$$\Omega_{2k,3} = A^3_{2k,3} = \{a_0, a_1, \ldots, a_{k+1}, \bar{a}_2, \bar{a}_3\}.$$

**Definition 1.12** *Let* $q = 2k, k \geq 4, p = 3$. *The substitution* $\sigma_{2k,3} : \Omega_{2k,3} \longrightarrow \Omega^*_{2k,3}$ *is defined as follows.*

$$\sigma_{2k,3}(a_0) = (a_1)^3;$$

$$\sigma_{2k,3}(a_i) = \bar{a}_2a_{i+1},\ 1 \leq i \leq k-2;$$

$$\sigma_{2k,3}(a_{k-1}) = \bar{a}_2;$$

$$\sigma_{2k,3}(a_k) = a_{k+1};$$

$$\sigma_{2k,3}(a_{k+1}) = \bar{a}_3a_3;$$

$$\sigma_{2k,3}(\bar{a}_j) = \bar{a}_{j+1}a_2,\ 2 \leq i \leq k-2;$$

$$\sigma_{2k,p}(\bar{a}_{k-1}) = a_ka_2.$$

**Case** $q = 6, k = 3, p = 3$

In this case

$$\Omega_{6,3} = A^3_{6,3} = \{a_0, a_1, a_2, a_3, a_4, \bar{a}_2\}.$$

**Definition 1.13** *Let* $q = 6, k = 3, p = 3$. *The substitution* $\sigma_{6,3} : \Omega_{6,3} \longrightarrow \Omega^*_{6,3}$ *is defined as follows:*

$$\sigma_{6,3}(a_0) = (a_1)^3;$$

$$\sigma_{6,3}(a_1) = \bar{a}_2 a_2;$$

$$\sigma_{6,3}(a_2) = \bar{a}_2;$$

$$\sigma_{6,3}(a_3) = a_4;$$

$$\sigma_{6,3}(a_4) = a_3;$$

$$\sigma_{6,3}(\bar{a}_2) = a_3 a_2.$$

**Case** $q = 2k + 1,\ k \geq 2,\ p \geq 4$

In this case

$$\Omega_{2k+1,p} = A^p_{2k+1,p} = \{a_0, a_1, \ldots, a_k, \bar{a}_1, \bar{a}_2, \ldots, \bar{a}_{k-1}, b_1, \ldots, b_k, \bar{b}_2, \bar{b}_3, \ldots, \bar{b}_k\}.$$

Here we use the notations of Lemmas 1.11 and 1.16.

**Definition 1.14** *Let* $q = 2k + 1,\ k \geq 2,\ p \geq 4$. *The substitution* $\sigma_{2k+1,p} : \Omega_{2k+1,p} \longrightarrow \Omega^*_{2k+1,p}$ *is defined as follows.*

$$\sigma_{2k+1,p}(a_0) = (b_1)^p;$$

$$\sigma_{2k+1,p}(a_i) = \bar{b}_2 (b_1)^{p-3} a_{i+1},\ 1 \leq i \leq k-1;$$

$$\sigma_{2k+1,p}(a_k) = \bar{b}_2 (b_1)^{p-4} b_2;$$

$$\sigma_{2k+1,p}(b_i) = \bar{b}_2 (b_1)^{p-3} b_{i+1},\ 1 \leq i \leq k-1;$$

$$\sigma_{2k+1,p}(b_k) = \bar{b}_2 (b_1)^{p-4} a_1;$$

$$\sigma_{2k+1,p}(\bar{a}_j) = \bar{a}_{j+1} (b_1)^{p-3} b_2,\ 1 \leq j \leq k-2;$$

$$\sigma_{2k+1,p}(\bar{a}_{k-1}) = (b_1)^{p-3} b_2;$$

$$\sigma_{2k+1,p}(\bar{b}_j) = \bar{b}_{j+1} (b_1)^{p-3} b_2,\ 2 \leq j \leq k-1;$$

$$\sigma_{2k+1,p}(\bar{b}_k) = \bar{a}_1 (b_1)^{p-4} b_2.$$

**Case** $q = 2k+1,\ k \geq 3,\ p = 3$

In this case

$$\Omega_{2k+1,3} = A^3_{2k+1,3} = \{a_0, a_1, \dots, a_k, a_{k+1}, \bar{a}_1, \bar{a}_2, \dots, \bar{a}_{k-1}, b_1, \dots, b_k, \bar{b}_2, \bar{b}_3, \dots, \bar{b}_k\}.$$

Here we use the notations of Lemma 1.11.

**Definition 1.15** *Let* $q = 2k+1, k \geq 3, p = 3$. *The substitution* $\sigma_{2k+1,3} : \Omega_{2k+1,3} \longrightarrow \Omega^*_{2k+1,3}$ *is defined as follows.*

$$\sigma_{2k+1,3}(a_0) = (b_1)^3;$$

$$\sigma_{2k+1,3}(a_1) = \bar{b}_3 a_2$$

$$\sigma_{2k+1,3}(a_i) = \bar{b}_2 a_{i+1},\ 2 \leq i \leq k-1;$$

$$\sigma_{2k+1,3}(a_k) = a_{k+1};$$

$$\sigma_{2k+1,3}(a_{k+1}) = \bar{b}_3 b_3;$$

$$\sigma_{2k+1,3}(b_i) = \bar{b}_2 b_{i+1},\ 1 \leq i \leq k-1;$$

$$\sigma_{2k+1,3}(b_k) = a_1;$$

$$\sigma_{2k+1,3}(\bar{a}_1) = \bar{a}_2 b_3;$$

$$\sigma_{2k+1,3}(\bar{a}_j) = \bar{a}_{j+1} b_2,\ 2 \leq j \leq k-2;$$

$$\sigma_{2k+1,3}(\bar{a}_{k-1}) = b_2;$$

$$\sigma_{2k+1,3}(\bar{b}_j) = \bar{b}_{j+1} b_2,\ 2 \leq j \leq k-1;$$

$$\sigma_{2k+1,3}(\bar{b}_k) = \bar{a}_1.$$

**Case** $q = 3,\ p \geq 6$

In this case

$$\Omega_{3,p} = A^p_{2k+1,p} = \{a_0, a, b\}.$$

We use the notations of Lemma 1.18

**Definition 1.16** *Let* $v \in V_{3,p}$, $p \geq 6$. *The substitution* $\sigma_{3,p} : \Omega_{3,p} \longrightarrow \Omega^*_{3,p}$ *is defined as follows.*

$$\sigma_{3,p}(a_0) = b^p;$$

$$\sigma_{3,p}(a) = a^{p-5} b;$$

$$\sigma_{3,p}(b) = a^{p-6} b.$$

**Theorem 1.1** *The tree $T_{q,p}$ corresponding to the orbit $(\sigma^n_{q,p}(a_0))_{n\geq 0}$ of $a_0$ with respect to the substitution $\sigma_{q,p} : \Omega_{q,p} \longrightarrow \Omega^*_{q,p}$ is a spanning tree of graph $\Gamma_{q,p}$. The vertices of $\Gamma_{q,p}$ of generation $n$ correspond to the letters of word $\sigma^n_{q,p}(a_0) \in \Omega_{q,p}$.*

Proof. Follows from the construction.

### 1.5 The transition graph, the transition matrix of substitution $\tilde{\sigma}_{q,p}$

Here we consider the substitution $\tilde{\sigma}_{q,p} = \sigma_{q,p}\Big|\tilde{\Omega}_{q,p} : \tilde{\Omega}_{q,p} \longrightarrow \tilde{\Omega}^*_{q,p}$, where $\tilde{\Omega}_{q,p} = \Omega_{q,p} \setminus \{a_0\}$.

Denote the transition(incidence) matrix of substitution $\tilde{\sigma}_{q,p}$ by $M_{q,p}$. The rows and the columns of $M_{q,p}$ are labeled by the elements $a_1, a_2, \ldots, b_1, b_2, \ldots, \bar{a_1}, \bar{a_2}, \ldots, \bar{b_2}, \bar{b_3}, \ldots$ of $\tilde{\Omega}_{q,p}$ in this order. The elements $m_{\alpha,\beta} \in M_{q,p}, \alpha, \beta \in \tilde{\Omega}_{q,p}$ are defined as

$$m_{\alpha,\beta} = card\{\beta \ : \ \beta \text{ is a letter of word } \tilde{\sigma}_{q,p}(\alpha)\}.$$

We denote by $\Gamma(\tilde{\sigma}_{q,p})$ the transition graph of substitution $\tilde{\sigma}_{q,p} : \tilde{\Omega}_{q,p} \longrightarrow \tilde{\Omega}^*_{q,p}$. Graph $\Gamma(\tilde{\sigma}_{q,p})$ is a directed graph with transition matrix $M_{q,p}$, [4], ch. 15, [1], ch. 3. A directed graph is called strongly connected if, for each pair of vertices $v_1$ and $v_2$, there is a directed path connecting $v_1$ to $v_2$.

Remind that the index of imprimitivity $h_{q,p}$ of the strongly connected graph $\Gamma(\tilde{\sigma}_{q,p})$ is the greatest common divisor of the lengths of the closed directed paths on $\Gamma(\tilde{\sigma}_{q,p})$, [1], ch. 3, 3.5, [4], ch. 15, 15.6.

**Lemma 1.19** *The transition graph $\Gamma(\tilde{\sigma}_{q,p})$ has the following properties:*

1. *$\Gamma(\tilde{\sigma}_{q,p})$ is strongly connected for $q = 2k, k \geq 3, p \geq 4$. The index of imprimitivity$h_{2k,p}$ of $\Gamma(\tilde{\sigma}_{q,p})$ is one;*
2. *$\Gamma(\tilde{\sigma}_{4,4})$ has two strongly connected components $\{a_1\}$ and $\{a_2\}$;*
3. *$\Gamma(\tilde{\sigma}_{2k,3})$, $k \geq 4$ has one strongly connected components. It contains all vertices different from $a_1$. The vertex $a_1$ is isolated. The index of imprimitivity $h_{2k,3}$ of the strongly connected component is one;*
4. *$\Gamma(\tilde{\sigma}_{6,3})$ has two strongly connected components $\{a_2, \bar{a_2}\}$, $\{a_3, a_4\}$. The vertex $a_1$ is isolated;*
5. *$\Gamma(\tilde{\sigma}_{2k+1,p})$, $k \geq 2$, $p \geq 4$ is strongly connected. The index of imprimitivity $h_{2k+1,p}$ of $\Gamma(\tilde{\sigma}_{2k+1,p})$ is one;*
6. *$\Gamma(\tilde{\sigma}_{2k+1,3})$, $k \geq 3$ has one strongly connected component, containing all vertices different from $b_1$. The index of imprimitivity $h_{2k+1,3}$ of this component is one;*
7. *$\Gamma(\tilde{\sigma}_{3,p})$, $p \geq 7$ is strongly connected. The index of imprimitivity $h_{3,p}$ of $\Gamma(\tilde{\sigma}_{3,p})$ is one;*
8. *$\Gamma(\tilde{\sigma}_{3,6})$ has two strongly connected components $\{a\}$ and $\{b\}$.*

Remind that the transition matrix of a strongly connected graph is irreducible, [4], ch. 15. Let $M$ be a nonnegative irreducible matrix, then the Perron-Frobenius theorem gives.:

1. Matrix $M$ has a positive eigenvalue, $r$, equal to the spectral radius of $M$;
2. There is a strongly positive (right) eigenvector associated with eigenvalue $r$;

3. Eigenvalue $r$ is a simple root of the characteristic polynomial of $M$. If $M = (m_{i,j})$ and $\sigma_j = \sum_k m_{j,k}$, then $min_j\, \sigma_j \leq r \leq max_j\, \sigma_j$;

4. Let the index of imprimitivity $h = h(M) > 1$. Then matrix $M$ has eigenvalues $\lambda_0 = r, \lambda_2, \ldots, \lambda_{h-1}$ with $\lambda_s = \exp \frac{2s\pi}{h} r, 0 \leq s \leq h-1$. All these eigenvalues are simple;

5. The index of imprimitivity $h = h(M)$ of matrix $M$ is equal to the index of imprimitivity of the transition graph of matrix $M$;

6. Matrix $M$ is called primitive if $h(M) = 1$. A primitive matrix $M$ has a simple positive eigenvalue $r$, which is dominant, i.e., $r > \lambda$ for every eigenvalue $\lambda \neq r$ of $M$, see [4], Ch. 15; [1], ch. 3.

**Corollary 1.2** *The transition matrix $M_{p,q}$ of substitution $\tilde{\sigma}_{q,p}$ has the following properties.*

1. $M_{2k,p},\ k \geq 3,\ p \geq 4$ *is primitive and it has a dominant single positive eigenvalue* $r_{2k,p} > 1$*;*

2. $M_{4,4}$ *is reducible;*

3. $M_{2k,3},\ k \geq 4$ *is reducible, but it has a dominant single positive eigenvalue* $r_{2k,3} > 1$*;*

4. $M_{6,3}$ *is reducible;*

5. $M_{2k+1,p},\ k \geq 2,\ p \geq 4$ *is primitive and it has a dominant single positive eigenvalue* $r_{2k+1,p} > 1$*;*

6. $M_{2k+1,3},\ k \geq 3$ *is matrix and it has a dominant single positive eigenvalue* $r_{2k+1,3} > 1$*;*

7. $M_{3,p},\ p \geq 7$ *is primitive and it has a dominant single positive eigenvalue* $r_{3,p} > 1$*;*

8. $M_{3,6}$ *is reducible.*

### 1.6 Characteristic polynomial of substitution $\tilde{\sigma}_{q,p}$ and number of elements of generation $n$ of $\Gamma_{q,p}$

By a definition, the characteristic polynomial $\chi_{q,p}(x)$ of substitution $\tilde{\sigma}_{q,p}$ is the characteristic polynomial of the matrix $M_{q,p}$. Let

$$\chi_{q,p}(x) = x^N - c_1 x^{N-1} - \cdots - c_{N-l-1}x - c_{N-l}x^l, l \geq 0$$

$$N = card(\tilde{\Omega}_{q,p}) = card(\Omega_{q,p}) - 1.$$

Let $u_n$ be the number of elements of generation $n$ of graph $\Gamma_{q,p}$. Observe that $u_n$ is the number of tiles of generation $n$ of tiling $\{p, q\}$. Theorem 1 implies that $u_n = card(\sigma^n(a_0)),\ n = 0, 1, \ldots$.

Denote by $\tilde{u}_n$ the number of elements of $\tilde{\sigma}^n(a)$, where $\sigma_{q,p}(a_0) = a^p, a \in \tilde{\Omega}_{q,p}$. Then $u_{n+1} = p\tilde{u}_n,\ n = 0, 1, \ldots$. Furthermore,

$$\tilde{u}_n = (1, 0, \ldots, 0)^T M^n_{q,p}(1, \ldots, 1),\ n = 0, 1, \ldots$$

Remind that the rows and columns of matrix $M_{q,p}$ are labeled by

$$a_1, a_2, \ldots, b_1, b_2, \ldots, \bar{a_1}, \bar{a_2}, \ldots, \bar{b_1}, \bar{b_2}, \ldots$$

in this order. The Cayley-Hamilton theorem gives that $\chi_{q,p}(M_{q,p}) = 0$. This implies that the sequence $\tilde{u}_n$ satisfies the recurrence

$$\tilde{u}_{n+N-l} = c_1\tilde{u}_{n+N-1} + c_2\tilde{u}_{n+N-2} + \cdots + c_{n-1}\tilde{u}_{n+1} + c_{N-l}\tilde{u}_n.$$

Sequence $\tilde{u}_n$ is determined by $\tilde{u_0}, \tilde{u_1}, \ldots, u_{N\tilde{-}l-1}$. They are given by $\tilde{u_j} = (1, 0, \ldots, 0)^T M^j_{q,p}(1, \ldots, 1),\ j = 0, 1, \ldots, N-l-1$.

Let $\lambda_1, \lambda_2, \ldots, \lambda_K$ be the roots of the characteristic polynomial $\chi_{q,p}(x)$ and assume that $\lambda_j$ has a multiplicity $d_j$, $j = 1, \ldots, K$, $d_1 + d_2 + \cdots + d_K = N$. We also assume that $|\lambda_1| \geq |\lambda_2| \geq \cdots \geq |\lambda_K|$. Then there exit constants (possible complex numbers) $\{C_{s,t_s} : s = 1, \ldots, K,\ t_s = 0, \ldots, d_s - 1\}$ such that

$$\tilde{u}_n = \sum_j^K (C_{j,0} + C_{j,1}n + C_{j,2}n^2 + \cdots + C_{j,d_j-1}n^{d_j-1})\lambda_j^n,$$

see [3].

Then $\tilde{u}_n$ grows as $|\lambda_1|^n$ in the case $|\lambda_1| > |\lambda_j|, j \neq 1$. Let us consider this question with more details.

**Theorem 1.2** *Let $u_n$ be the number of the elements of generation $n$ of graph $\Gamma_{q,p}$ and let $u_{n+1} = \tilde{u}_n$, $n = 0, 1, \ldots$.*

*Then $\tilde{u}_n \approx r^n_{q,p}$ in the cases*

1. *(a)* $q = 2k,\ k \geq 3,\ p \geq 4$;
   *(b)* $q = 2k,\ k \geq 4,\ p = 3$;
   *(c)* $q = 2k+1,\ k \geq 2,\ p \geq 4$;
   *(d)* $q = 2k+1,\ k \geq 3,\ p = 3$;
   *(e)* $q = 3,\ p \geq 7$.

   *Moreover $p - 2 < r_{q,p} < p - 1$.*

   *In these cases the growth is exponential.*

2. *(a)* $\chi_{4,4} = (x-1)^2$. *Then* $\tilde{u}_n \approx n$;
   *(b)* $\chi_{6,3} = -x^3(x^2-1)$. *Then* $\tilde{u}_n \approx n$;
   *(c)* $\chi_{3,6} = (x-1)^2$. *Then* $\tilde{u}_n \approx n$.
   *In these cases the growth is linear.*

Calculations (direct or with Maple 8) give us an evidence that the characteristic polynomial $\chi_{q,p}(x)$is

- $$\chi_{2k,p}(x) = (x^{k-2} + x^{k-1} + \cdots + x + 1)(x^k - (p-2)x^{k-1} - \cdots - (p-2)x + 1),$$
  for $k \geq 3,\ p \geq 4$;

- $$\chi_{2k+1,p}(x) = (x^{2k-2} + x^2k + \cdots + x + 1)(x^{2k} - (p-2)x^{2k-1} - \cdots - (p-2)x^{k+1} - (p-4)x^k - (p-2)x^{k-1} - \ldots - (p-2)x + 1),$$
  for $k \geq 3,\ p \geq 4$;

- $$\chi_{2k,3}(x) = -x(x^{k-2} + x^{k-1} + \cdots + x + 1)(x^k - x^{k-1} - \cdots - x + 1),$$
  for $k \geq 4$;

- $$\chi_{2k+1,3}(x) = \chi_{q,p}(x) = -x(x^{2k-2} + x^{2k} + \cdots + x + 1)(x^{2k} - x^{2k-1} - \cdots - x^{k+1}x^{k+1} + x^k - x^{k-1} - \cdots - x + 1),$$
  for $k \geq 3$;

- $$\chi_{3,p}(x) = x^2 - (p-4)x + 1,$$
  for $p \geq 7$;

The characteristic polynomial $\chi_{2q,p}(x)$ has a nontrivial factor

$$\chi_{q,p} = x^k - (p-2)x^{k-1} - \cdots - (p-2)x + 1$$

in the cases $q = 2k,\ k \geq 3,\ p \geq 4$ and $q = 2k,\ k \geq 4, p = 3$. We have numerical evidence (based on calculations with Maple 8) that all zeros of this factor different from $r_{2k,p}$ and $r_{2k,p}^{-1}$ have absolute value one, i.e., they are points of the unite circle.

The characteristic polynomial $\chi_{2k+1,p}(x)$ has a nontrivial factor

$$\chi_{2k+1,p}(x) = (x^{2k} - (p-2)x^{2k-1} - \cdots - (p-2)x^{k+1} - (p-4)x^k - (p-2)x^{k-1} - \cdots - (p-2)x + 1)$$

in the cases $q = 2k+1,\ k \geq 2,\ p \geq 4$ and $q = 2k+1,\ k \geq 3, p = 3$. We have numerical evidence (based on calculations with Maple 8) that all zeros of this factor different from $r_{2k+1,p}$ and $r_{2k+1p}^{-1}$ have absolute value one, i.e. they are points of the unite circle.

**Remark** When $q = 4k+1$, as polynomial $\chi_{4k+1,p}(x)$ is reciprocal, it is not difficult to prove that it if it has exactly two real roots, then it has at least one complex root $z_1$ such that $|z_1| = 1$. Indeed, if $z_1$ is a complex root with $|z_1| \neq 1$, then, $z_1$, $\overline{z_1}$, $\frac{1}{z_1}$ and $\frac{1}{\overline{z_1}}$ are four distinct roots. Accordingly, if no complex root has modulous 1, and if the polynomial has exactly two real roots, then the number of roots is of the form $4h$+2.

### 1.7 Acknowledgement

The authors thank the University of Bremen and the University of Metz for helping them to maintain the fruitful contacts which allowed them to write this paper.